\documentclass[preprints,article,accept,pdftex,moreauthors]{Definitions/mdpi} 

\usepackage{siunitx}
\def\simlt{\mathrel{\rlap{\lower 3pt\hbox{$\sim$}}\raise 2.0pt\hbox{$<$}}}    
\def\simgt{\mathrel{\rlap{\lower 3pt\hbox{$\sim$}} \raise 2.0pt\hbox{$>$}}}   
\firstpage{1} 
\pubvolume{1}
\issuenum{1}
\articlenumber{0}
\pubyear{2024}
\copyrightyear{2024}
\externaleditor{Academic Editor: Oleg Malkov }
\datereceived{2024-05-22 } 
\daterevised{ 2024-07-01} 
\dateaccepted{ 2024-07-03} 
\datepublished{ } 
\hreflink{https://doi.org/} 

\Title{Correlations between IR 
 Luminosity, Star Formation Rate, and CO Luminosity in the Local Universe} 

\TitleCitation{Correlations between IR Luminosity, Star Formation Rate, and CO Luminosity in the Local Universe}

\Author{Matteo 
 Bonato $^{1,2}$*\orcidA{}, Ivano Baronchelli $^{1,2}$, Viviana Casasola $^{1}$, Gianfranco De Zotti $^{3}$, Leonardo Trobbiani $^{1,4}$, Erlis Ruli $^{5}$, Vidhi Tailor $^{1,4}$ and Simone Bianchi $^{6}$}

\AuthorNames{Matteo Bonato, Ivano Baronchelli, Viviana Casasola, Gianfranco De Zotti, Leonardo Trobbiani, Erlis Ruli, Vidhi Tailor and Simone Bianchi}

\AuthorCitation{Bonato, 
 M.; Baronchelli, I.; Casasola, V.; De Zotti, G.; Trobbiani, L.; Ruli, E.; Tailor, V.; Bianchi, S.}

\address{%
$^{1}$ \quad  INAF---Istituto 
 di Radioastronomia, Via Gobetti 101, I-40129 Bologna, Italy; 
\\
$^{2}$ \quad  Italian ALMA Regional Centre, Via Gobetti 101, I-40129 Bologna, Italy;\\
$^{3}$ \quad  INAF---Osservatorio Astronomico di Padova, Vicolo dell’Osservatorio 5, I-35122 Padova, Italy;\\
$^{4}$ \quad  Dipartimento di Fisica e Astronomia, Alma Mater Studiorum Universit\`a di Bologna, Via Piero Gobetti 93/2, I-40129 Bologna, Italy;\\
$^{5}$ \quad  Dipartimento di Scienze Statistiche, Universit\`a degli Studi di Padova, Via Cesare Battisti 241/243, \mbox{I-35121 Padova, Italy};\\
$^{6}$ \quad INAF---Osservatorio Astrofisico di Arcetri, Largo E. Fermi 5, 50125 Firenze, Italy.}

\corres{Correspondence: matteo.bonato@inaf.it}

\abstract{We exploit the DustPedia sample of galaxies within approximately 40 Mpc, selecting 
 388~sources,  to investigate the correlations between IR luminosity (L$_{\rm IR}$), the star formation rate (SFR), and the CO(1-0) 
 luminosity (L$_{\rm CO}$) down to much lower luminosities than reached by previous analyses. We find a sub-linear dependence of the SFR on  L$_{\rm IR}$.  Below $\log(\hbox{L}_{\rm IR}/\hbox{L}_\odot)\simeq 10$ or \mbox{$\hbox{SFR}\simeq 1\,\hbox{M}_\odot\,\hbox{yr}^{-1}$}, the SFR/L$_{\rm IR}$ ratio substantially exceeds the standard ratio for dust-enshrouded star formation, and the difference increases with decreasing L$_{\rm IR}$ values. This implies that the effect of unobscured star formation overcomes that of dust heating by old stars, at variance with results based on the \textit{Planck} 
 ERCSC galaxy sample. We also find that the relations between the L$_{\rm CO}$ and L$_{\rm IR}$ or the SFR are consistent with those obtained at much higher luminosities. }

\begin{document}


\section{Introduction}\label{sect:intro}


When the light emitted by young stellar populations is absorbed and re-emitted by dust, the total (8--$1000\,\upmu$m, rest frame) infrared (IR) luminosity, L$_{\rm IR}$, is an effective measure of the star formation rate (SFR; 
\cite{Kennicutt1998}). These conditions are generally present in high-redshift, IR-luminous, dusty star-forming galaxies and in dense circum-nuclear starbursts at lower redshifts. However, in low-$z$ disk and ``normal'' early-type galaxies the situation is more complicated. On the one hand, blue star-formation regions show that the light from young stars is at least partly unobscured. On the other hand, the IR emission shows a cold ``cirrus'' contribution, powered by the general interstellar radiation field, dominated by the old stellar population. The ``cirrus'' component may prevail especially in early-type galaxies. The two effects work in opposite directions and the result of their combination is largely unknown.

Dust luminosity has also been found to be a useful tracer of the molecular gas mass, alternative to the most commonly used CO(1-0) line (e.g., \cite{Dunne2022}, and references therein). Molecular gas is the star-formation fuel; hence, it has a key role in galaxy evolution models. This has motivated investigations of correlations between the L$_{\rm IR}$ or the SFR and the L$_{\rm CO}$; however, these are limited to relatively high IR luminosity.

The complications mentioned above are particularly conspicuous at low IR luminosity/SFR. A preliminary investigation of the \mbox{L$_{\rm IR}$--SFR} correlation in  this regime was carried out by \mbox{\citet{Clemens2013}}.  They used a sample of 234 galaxies with flux density greater than 1.8\,Jy at 545\,GHz, located within approximately 100\,Mpc (\citet{Negrello2013}), drawn from the Planck Early Release Compact Source Catalogue (ERCSC; \citep{ERCSC}). The distribution of their infrared luminosities peaked at $\log(\hbox{L}_{\rm IR}/\hbox{L}_\odot) = 10$ with tails extending over the range   $7.5< \log(\hbox{L}_{\rm IR}/\hbox{L}_\odot) < 12$.

In this paper, we exploit the DustPedia database\endnote{\url{http://dustpedia.astro.noa.gr/}}, 
 which allows us to explore the \mbox{L$_{\rm IR}$--SFR} relation  down to the  dust luminosity much fainter than that reached by the sample used by \mbox{\citet{Clemens2013}}. It also allows us to investigate the correlation between L$_{\rm IR}$ and L$_{\rm CO}$ in this luminosity regime. The galaxies in the sample are morphologically classified so that it is be possible to look for similarities and differences between early- and late-type~populations.

The rest of this paper is structured as follows. Section~\ref{sec:data} presents a short description of the sample. In Section~\ref{sec:results}, the L$_{\rm IR}$--SFR--L$_{\rm CO}$ correlations are analyzed and described. The main conclusions are presented in Section~\ref{sec:conclusions}.  

\section{Data}\label{sec:data}

DustPedia (\citet{Davies2017}) is a large collaborative project aimed at substantially improving our understanding of the properties of dust in nearby galaxies and of its influence on physical processes occurring in the interstellar medium.

The DustPedia sample includes galaxies within a recession velocity of approximately $3000\,$km/s (corresponding to a distance of about 40\,Mpc) and a major axis of the isophote at the optical surface brightness of $25\,\hbox{mag}\,\hbox{arcsec}^{-2}$, $D_{25} > 1^\prime$. Galaxies were further required to be detected by the \textit{Herschel} Space Telescope mission and to be brighter than the Wide-Field Infrared Survey Explore (WISE) $3.4\,\upmu$m all-sky $5\,\sigma$ sensitivity limit.

\textls[15]{The sample reaches $\log(\hbox{L}_{\rm IR}/\hbox{L}_\odot) \simlt 6.5$ and $\log(\hbox{SFR}/M_\odot\,\hbox{yr}^{-1})\simlt -3$. Luminosities in the CO(1–0) line (L$_{\rm CO}$) were collected from the literature and homogenized by \mbox{\citet{Casasola2020}} [see their Table A.1, 
 for the references of data sources]. The CO measurements are available for only $\sim$24\% of the sample and refer to late-type galaxies only. CO luminosities were computed using the distances in the DustPedia catalogue, which are redshift-independent in most cases and were otherwise computed using \linebreak  \mbox{$h=H_0/100\, \rm km\,s^{-1}\,Mpc^{-1} = 0.73$.}}

The photometric DustPedia dataset consists of 875 galaxies for which multi-wavelength aperture-matched photometry for up to 41 bands, from far UV- to microwaves, was performed uniformly  \citep{Clark2018}. \citet{Nersesian2019} dropped 61 galaxies with poor quality measurements in some bands or insufficient coverage of the spectral energy distribution (SED)  from this dataset. 

To the remaining 814 galaxies they applied the SED fitting Code Investigating GALaxy Emission (CIGALE; \citep{Burgarella2005,Noll2009,Boquien2019}), upgraded implementing The Heterogeneous Evolution Model for Interstellar Solids (THEMIS; \citep{Jones2013,Jones2017}) dust model. Many DustPedia galaxies also belong to the Updated Nearby Galaxy Catalog, for which \citet{Karachentsev2013} derived SFR estimates from H$\alpha$ and UV measurements. We checked that such estimates are in very good agreement with the CIGALE's ones. 

Through CIGALE, they derived the physical properties of each galaxy, including the stellar mass, the current SFR, the dust mass and temperature, and the total dust luminosity, L$_{\rm dust}$ (essentially equal to L$_{\rm IR}$), with their uncertainties. They also investigated the dust heating by the old (age $>$ 200 Myr) and young ($\leq$200 Myr) stellar populations separately. 

We further excluded the 19 galaxies for which the results could be biased by the presence of an AGN \citep{Bianchi2018}~(no AGN model was used in the fits) and the four galaxies which may have a substantial radio contribution to the far-infrared (FIR) to millimetre (mm) luminosity \citep{Nersesian2019}. We were then left with 791 galaxies. A large fraction of them have a weak FIR to mm emission, so their total dust luminosity and SFR are highly uncertain. To avoid an excessive spread of the distribution of the SFR versus L$_{\rm IR}$ due to uncertainties, we limited our analysis to galaxies for which the estimates of each of these quantities are at least three times their respective uncertainties. This left 388 sources which constitute our final sample. 

We consider two morphological classes, early-type galaxies (ETGs) and late-type galaxies (LTGs). We used the classification by \citet{Casasola2020}, which is based on de \mbox{Vaucouleurs' \citep{deVaucouleurs1991}} numerical type, T, complemented with additional criteria which affect a minor fraction of sources. The type T was taken from the HyperLEDA catalogue of redshift-independent extragalactic distances (\citet{Makarov2014}). ETGs include ellipticals and S0s, while LTGs comprise spirals and irregulars. 



A table with classification, L$_{\rm IR}$, SFR, S$_{\rm CO(1\text{-}0)}$, and 
 L$_{\rm CO}$ measurements for the sources of our final sample is provided as online Supplementary Material.

\begin{figure}[H] 
\includegraphics[width=0.325\textwidth]{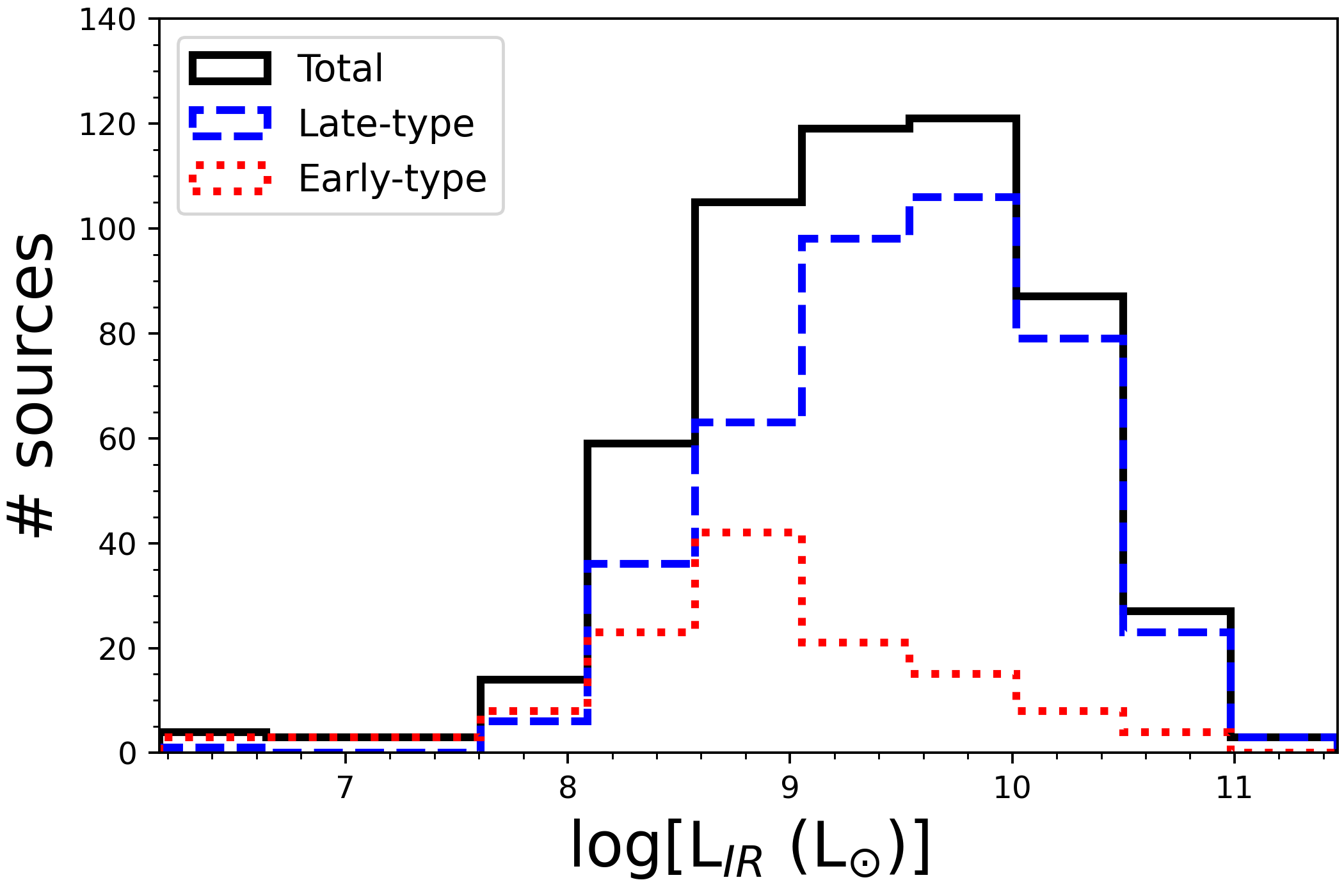}
\includegraphics[width=0.325\textwidth]{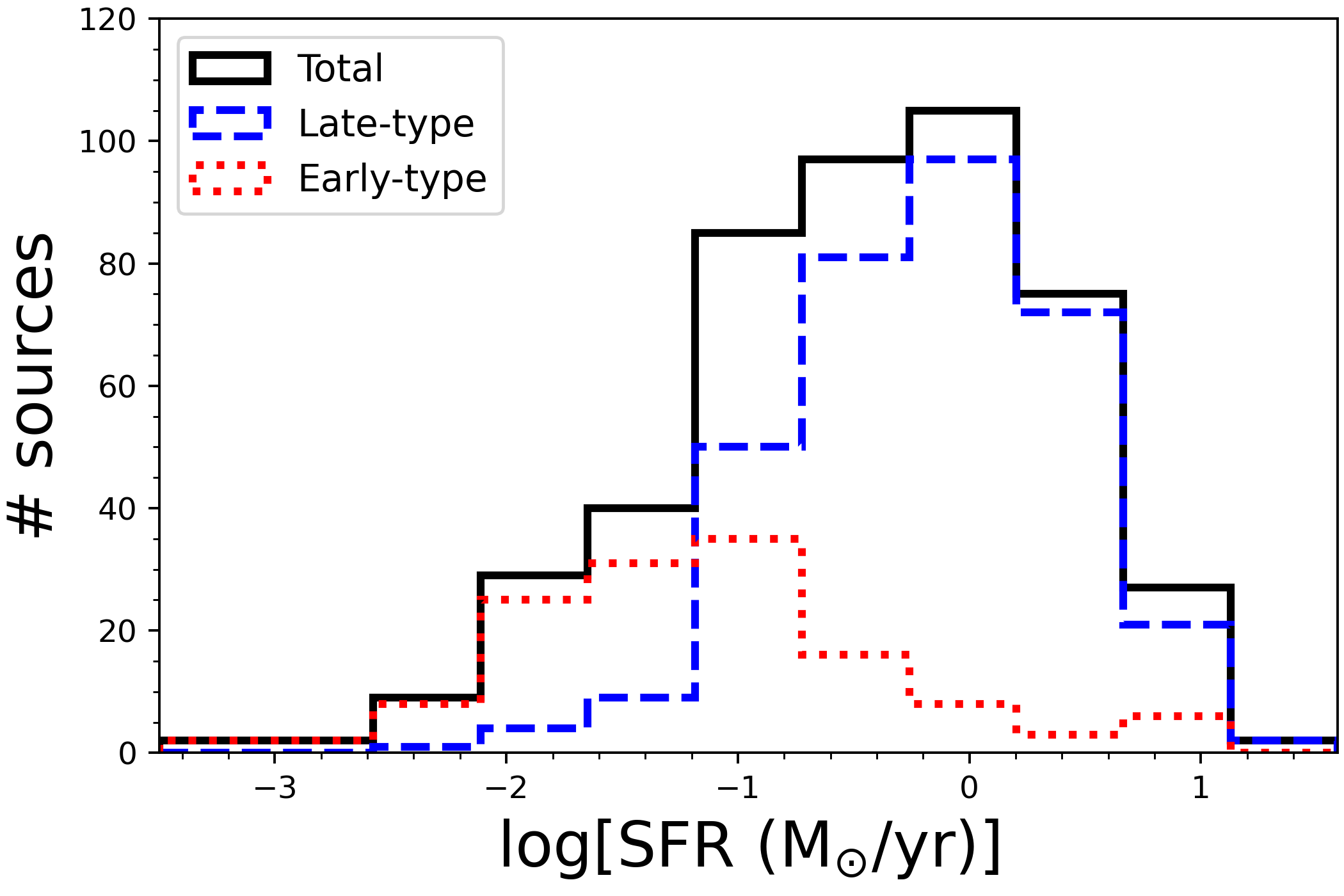}
\includegraphics[width=0.325\textwidth]{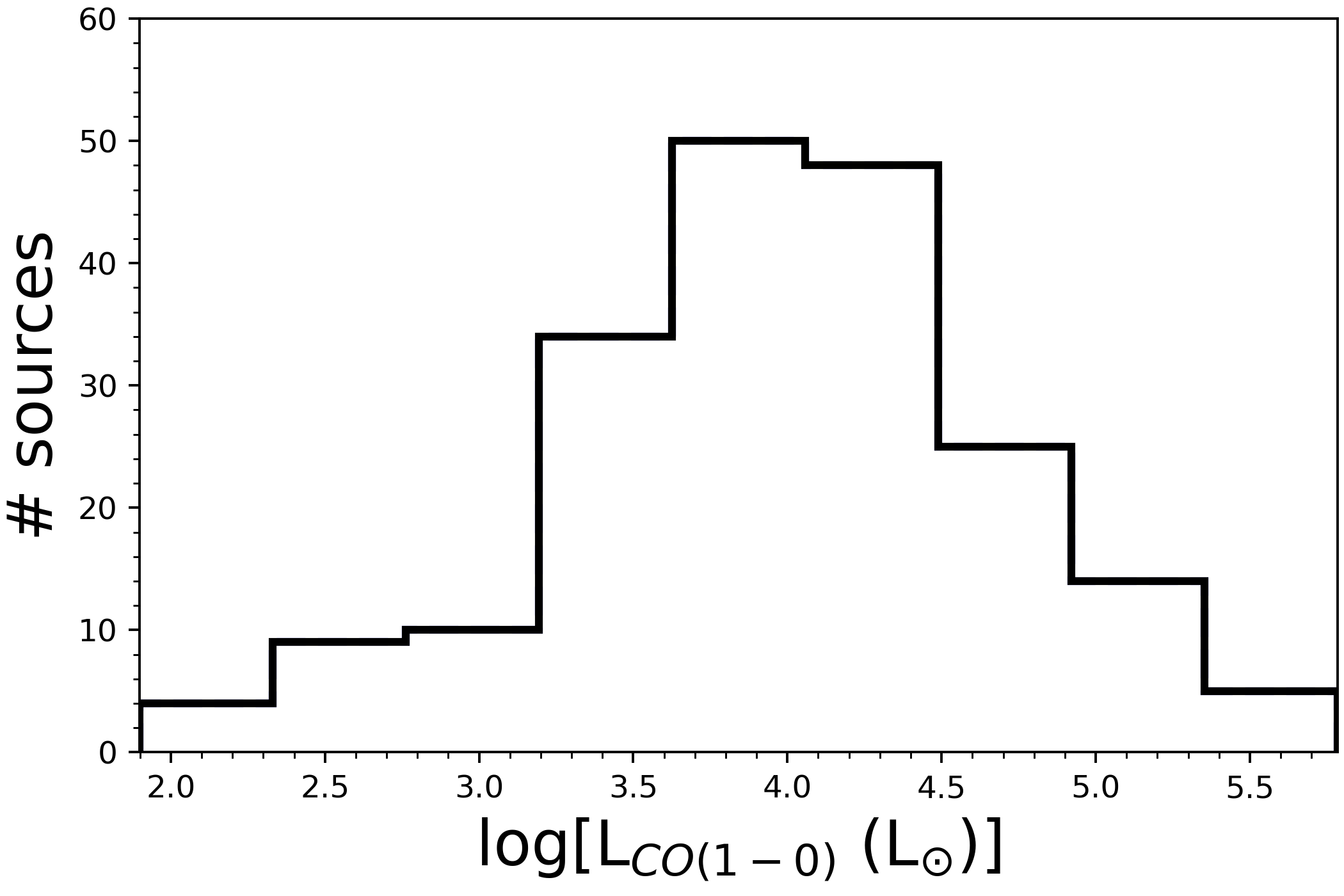}
\caption{Distributions 
 of $\log(\rm L_{\rm IR})$ (\textbf{left}), $\log(\rm SFR)$ (\textbf{center}), and $\log(\rm L_{\rm CO})$ (\textbf{right}) for our final sample: late-type galaxies (dashed blue lines), early-type galaxies (dotted red lines), and total (solid black lines). In the case of the distribution of $\log(\rm L_{\rm CO})$, only the total is shown; as mentioned in Section\,\protect\ref{sect:intro}, CO(1-0) measurements are available for a subset of late-type galaxies only.  }
\label{fig:distributions}
\end{figure}

The distributions of the L$_{\rm IR}$, SFR, and L$_{\rm CO}$ for our sample are shown in Figure\,\ref{fig:distributions}. The peaks of the distributions are as follows: $\rm L_{\rm IR}\sim$5.99 $\times$ 10$^{9}$ (for the total and the late-type populations), 6.50 $\times$ 10$^{8}$\,L$_{\odot}$ (for the early types); $\rm SFR\sim$0.94 (for the total and the late-type populations), 0.11\,M$_{\odot}$/yr (for early types); and $\rm L_{\rm CO}\sim$6.94 $\times$ 10$^{3}$\,L$_{\odot}$ (only a subsample of our late-type galaxies have CO measurements).\vspace{-2pt}

\vspace{-9pt}



\section{Results and Discussion}\label{sec:results}

\subsection{L$_{\rm IR}$--SFR Correlation}\label{subsec:results_1}



\begin{figure}[H] 
\includegraphics[width=0.325\textwidth]{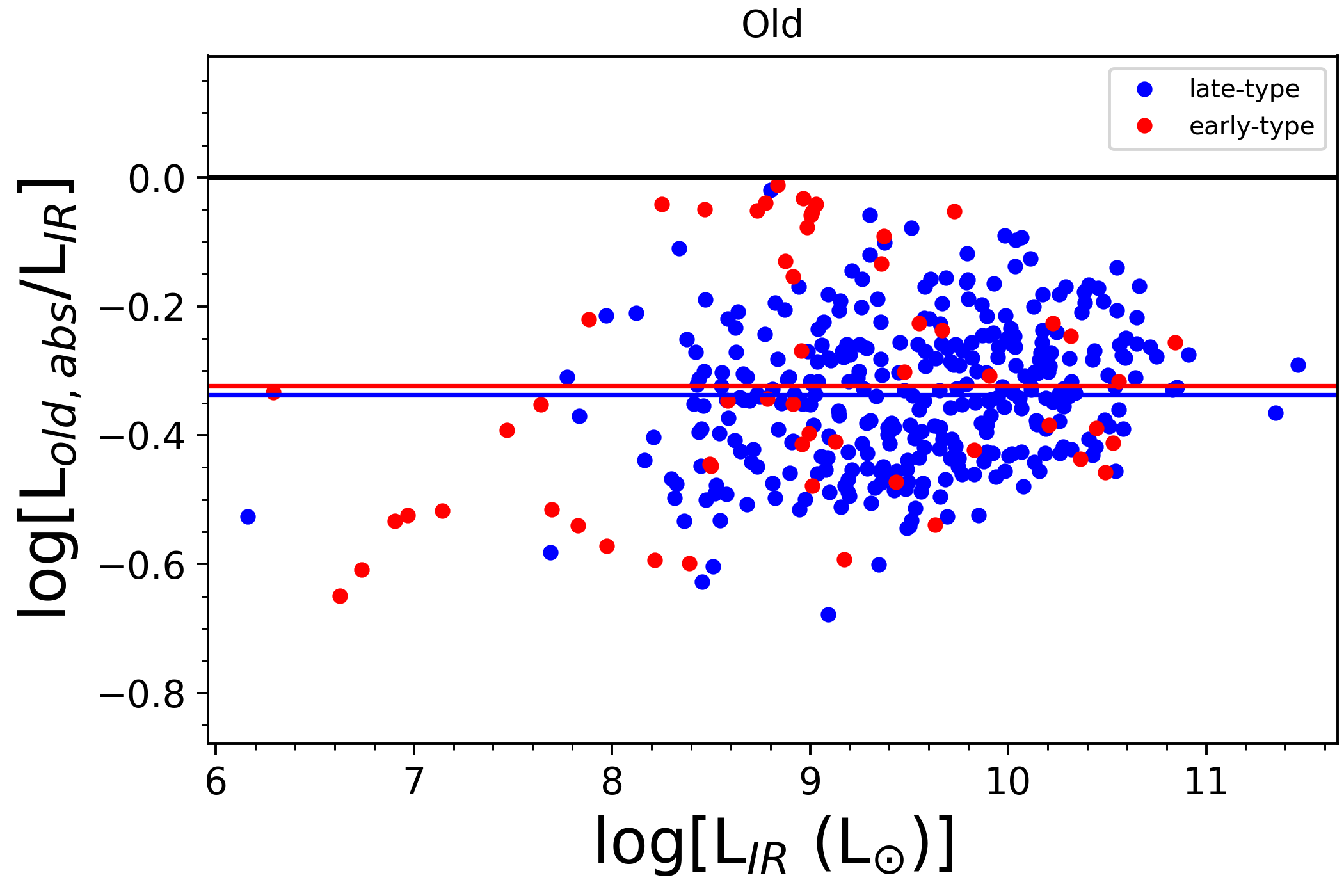}
\includegraphics[width=0.325\textwidth]{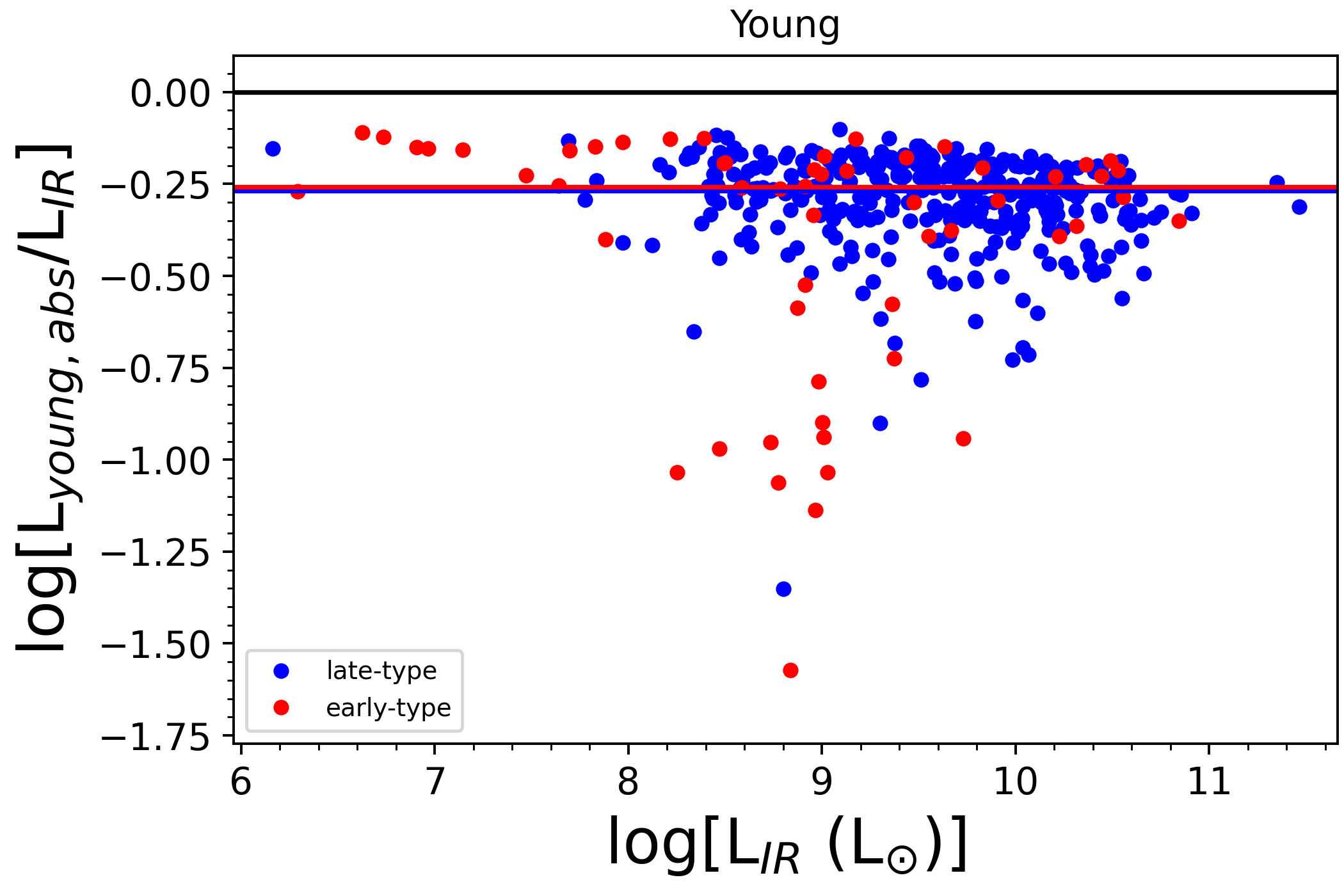}
\includegraphics[width=0.325\textwidth]{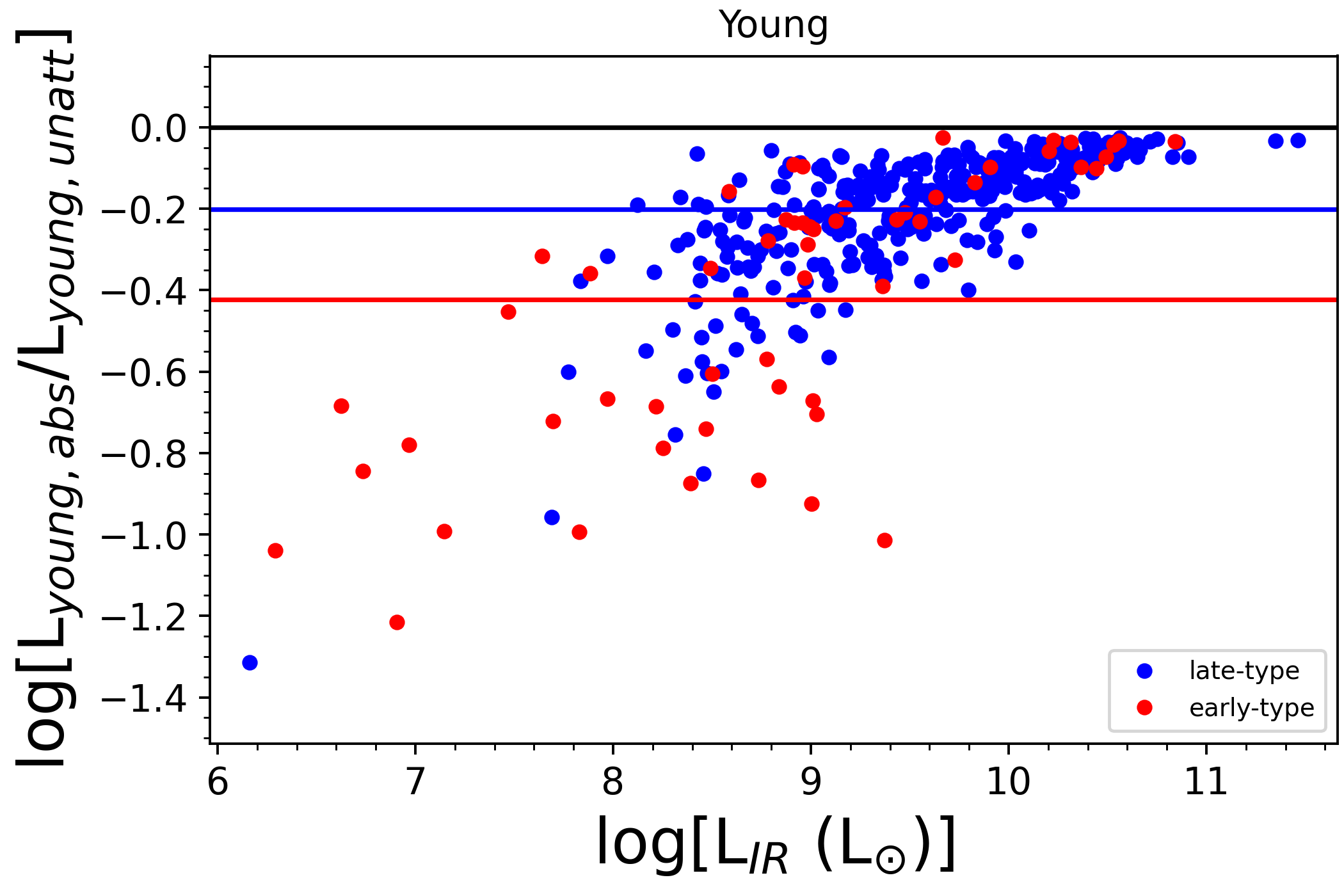}
\caption{Ratio 
 between the absorbed luminosity, L$_{\rm abs}$, of old (\textbf{left}) and young (\textbf{center}) stellar populations and the total dust emission luminosity, L$_{\rm IR}$. The right-hand panel shows the absorbed fraction of the starlight emitted by young stars, as a function of L$_{\rm IR}$, for early- and late-type galaxies. The blue and red dots represent the late- and early-type galaxies, respectively; the horizontal blue and red lines correspond to the average values for the two galaxy types. The horizontal black line corresponds to unit ratios.  We adopted the luminosities derived by \citet{Nersesian2019} using CIGALE SED~fitting. 
}
\label{fig:Ldust_vs_Lstar}
\end{figure}

The left and central panels of Figure\,\ref{fig:Ldust_vs_Lstar} show the contributions, represented by the absorbed\endnote{CIGALE, the  SED fitting code employed by \citet{Nersesian2019} on DustPedia galaxies, has the capability of disentangling attenuated and unattenuated emissions from old and young stellar populations. This enables us to quantify the fraction of energy absorbed by dust for each stellar component.} starlight luminosity, L$_{\rm abs}$, of old and young stellar populations to the total dust emission luminosity, L$_{\rm IR}$. In the case of old stellar populations, the mean values of $\log(\hbox{L}_{\rm abs}/\hbox{L}_{\rm IR})$ for LTGs and ETGs are very close to each other, \mbox{$\langle\log(\hbox{L}_{\rm abs}/\hbox{L}_{\rm IR})\rangle \simeq -0.34\pm0.01$} and $\simeq-$0.32 $\pm$ 0.02, respectively, corresponding to contributions of 46--48\%. In some extreme cases, mostly associated with ETGs, dust heating is almost entirely due to old~stars.\vspace{-2pt}

Furthermore, in the case of the young stellar populations the median value of $\log(\hbox{L}_{\rm abs}/\hbox{L}_{\rm IR})$ is comparable for LTGs and ETGs, $\simeq-0.27\pm0.01$ and $\simeq-0.26\pm0.06$, respectively, corresponding to contributions to L$_{\rm IR}$ of $\simeq 54\%$ and $\simeq 55\%$, respectively. At the lowest IR luminosities ($\log(\hbox{L}_{\rm IR}/\hbox{L}_\odot)\le 8$), mostly associated with ETGs, dust heating is almost entirely due to young stars.

As expected, in LTGs the average contribution to dust heating of the young stellar population is larger than that of old stars, but the contribution of the latter is substantial. 
The same behavior is also shown by ETGs.

The right-hand panel of Figure\,\ref{fig:Ldust_vs_Lstar} shows the dust-absorbed fraction of starlight emitted by young stellar populations in ETGs and LTGs as a function of the total IR luminosity. Such fraction is close to unity at high IR luminosity, but decreases, on average, with decreasing L$_{\rm IR}$ values. For $\log(\hbox{L}_{\rm IR}/\hbox{L}_\odot)\le 9$, more that half of the emission is unobscured, even for LTGs. The mean values of $\log(\hbox{L}_{\rm young, abs}/\hbox{L}_{\rm young, unatt})$ are $-0.20\pm0.01$ for LTGs and $-0.42\pm0.04$ for ETGs.

\begin{figure}[H] 
\includegraphics[width=0.9\textwidth]{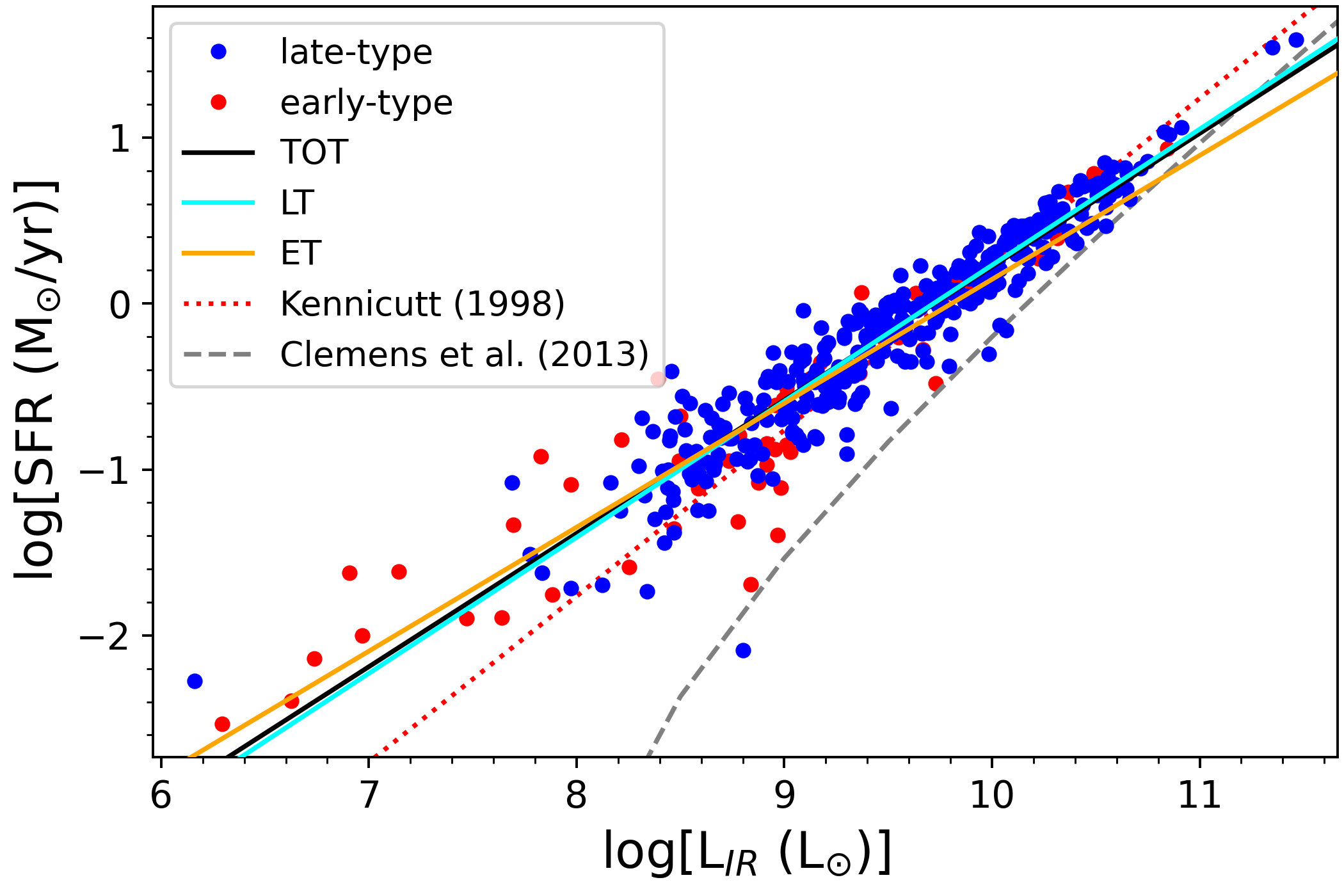}
\caption{SFR 
 versus L$_{\rm IR}$ for late- and early-type galaxies (blue and red dots, respectively). The solid cyan and orange lines show the best-fit relations for the two populations. The dotted red line shows the standard Kennicutt \cite{Kennicutt1998} relation. The dashed grey line shows the best-fit relation between SFR and \textit{total} $L_{\rm IR}$ derived by Clemens et al. \cite{Clemens2013} for their sample. The typical uncertainties have a size close to those of data points.}
\label{fig:SFRvsLdust}
\end{figure}

Figure\,\ref{fig:SFRvsLdust} shows $\log(\text{SFR})$ versus $\log(\rm L_{\rm IR})$ for both early- and late-type galaxies (red and blue dots, respectively). We modeled the relation among the two quantities by means of a robust linear regression methodology\endnote{Robust linear regression down-weights extreme data points to obtain a more stable fit.}. In particular, the approach used is that of a robust $M$-estimator\endnote{M-estimators are a class of estimators that extend the well-known maximum likelihood estimators (for further details, see Chapter 3 of \cite{huber2009robust}).}. This methodology is implemented in the function \textit{rlm} in the \textit{R} package \textit{MASS}.\vspace{-3pt}

In particular, we fitted a simple linear regression model,
\begin{linenomath}
\begin{equation}
\log (\text{SFR}_i) = \beta_0 + \beta_1 \log(\text{L}_{\rm IR,i}) + \epsilon_i,
\label{eq:0}
\end{equation}
\end{linenomath}
where $\epsilon_i$ is the error term assumed to be identically and independently $N(0, \sigma^2)$-distributed, $i = 1, \ldots, n.$

The result for the total sample is
\begin{linenomath}
\begin{equation}
\log (\rm SFR/M_{\odot}\,\hbox{yr}^{-1}) = (-7.81\pm0.11) + (0.80\pm0.01) \times \log (\rm L_{\rm IR}/L_{\odot}),
\label{eq:1_tot}
\end{equation}
\end{linenomath}
where the uncertainties are the estimated standard errors (the robust confidence intervals at 95\% confidence level for $\beta_0$ and $\beta_1$ are [$-$8.03,$-$7.60] and [0.78,0.83], respectively).

For the late-type population, we obtained the following relation:
\begin{linenomath}
\begin{equation}
\log (\rm SFR/M_{\odot}yr^{-1}) = (-7.97\pm0.13) + (0.82\pm0.01) \times \log (\rm L_{\rm IR}/L_{\odot}).
\label{eq:1_lt}
\end{equation}
\end{linenomath}

The 
 robust confidence intervals at a 95\% confidence level for the intercept and the slope parameters are [$-$8.21,$-$7.72] and [0.79,0.85], respectively.

For the early-type population, we obtained the following relation:
\begin{linenomath}
\begin{equation}
\log (\rm SFR/M_{\odot}yr^{-1}) = (-7.32\pm0.34) + (0.75\pm0.04) \times \log (\rm L_{\rm IR}/L_{\odot}),
\label{eq:1_et}
\end{equation}
\end{linenomath}

The robust confidence intervals at the 95\% confidence level for the intercept and the slope parameters are [$-$8.00,$-$6.65] and [0.67,0.82], respectively.

As illustrated by Figure\,\ref{fig:SFRvsLdust}, the best-fit relations for ETGs and LTGs are almost coincident. At medium to high luminosity, the relations agree with the standard Kennicutt \cite{Kennicutt1998} relation, which we chose in place of the more recent relation by Kennicutt and Evans \cite{Kennicutt2012} because it refers to the Salpeter initial mass function (IMF) also used for the DustPedia project\endnote{Please note that 
the Kennicutt and Evans \cite{Kennicutt2012} relation refers to the Kroupa IMF, which is very similar to the Chabrier IMF used by Clemens~et~al.~\cite{Clemens2013}. Instead, the Kennicutt \cite{Kennicutt1998} relation and the Dustpedia fits refer to the Salpeter IMF. However, the difference is only a factor of 0.86 (Table 1 
 of Kennicutt and Evans \cite{Kennicutt2012}), meaning that the Kennicutt and Evans \cite{Kennicutt2012} calibration is lower (in SFR) than the Kennicutt \cite{Kennicutt1998} one by this factor. In Figure\,\ref{fig:SFRvsLdust}, for an appropriate comparison we chose to use the Kennicutt  \cite{Kennicutt1998} relation and raise the Clemens~et~al.~\cite{Clemens2013} relation by a factor of $1/0.86$, which in practice brings everything to the Salpeter IMF. Alternatively, for an equally appropriate comparison, we could have lowered our SFRs by the factor of 0.86 and compared them to the Kennicutt and Evans \cite{Kennicutt2012} relation and the original Clemens~et~al.~\cite{Clemens2013}.}. 

The increase in $\log(\hbox{SFR})$ with $\log(\hbox{L}_{\rm IR})$ is sub-linear. At low luminosity, the SFRs estimated from CIGALE are substantially higher than expected from the Kennicutt  \cite{Kennicutt1998} relation, implying that a large fraction of the light from young stars is not absorbed by dust, as also shown by Figure\,\ref{fig:Ldust_vs_Lstar}. In other words, using the IR luminosity alone and the Kennicutt \cite{Kennicutt1998} relation leads to underestimation of the SFR; a correct estimate requires taking into account both the absorbed and the unabsorbed light from young stars.

The data points are consistently above the best-fit relation between the SFR and \textit{total} L$_{\rm IR}$, the sum of the contributions of young and old stellar populations, derived by Clemens~et~al.~\cite{Clemens2013} for their sample (dashed grey line). The different selection of the sample, the different resolution of the \textit{Herschel} photometry used by DustPedia compared to the \textit{Planck}'s used by Clemens~et~al.~\cite{Clemens2013}, 
 and the fact that these authors used the preliminary photometry contained in the \textit{Planck} ERCSC can have a role. We note, in particular, that ERCSC flux densities for extended galaxies, like those in the DustPedia sample, were found to be larger by substantial factors than the \textit{Herschel} ones in nearby bands (see, e.g., \cite{Herranz2013}).

\label{eq:2_lt}






\subsection{The L$_{\rm CO}$--L$_{\rm IR}$ and L$_{\rm CO}$--SFR Correlations}\label{subsec:results_2}

 As stated in Section\,\ref{sect:intro}, we only have CO(1$-$0) measurements (provided as online Supplementary Material) for a subset of LTGs. We used the robust linear regression methodology described in Section\,\ref{subsec:results_1} again to investigate the relation between L$_{\rm CO}$ and L$_{\rm IR}$. We~obtained the following: 
\begin{linenomath}
\begin{equation}
\log \rm L_{\rm CO} = (-5.75\pm0.47) + (0.99\pm0.05) \times \log \rm L_{\rm IR},
\label{eq:3}
\end{equation}
\end{linenomath}
where both luminosities are in units of L$_\odot$. The robust confidence intervals at a 95\% confidence level for the intercept and the slope are [$-$7.29,$-$4.22] and [0.84,1.15], respectively. 

\begin{figure}[H]
\text{~}\\	
\vspace{-10pt}	

\includegraphics[width=0.99\textwidth]{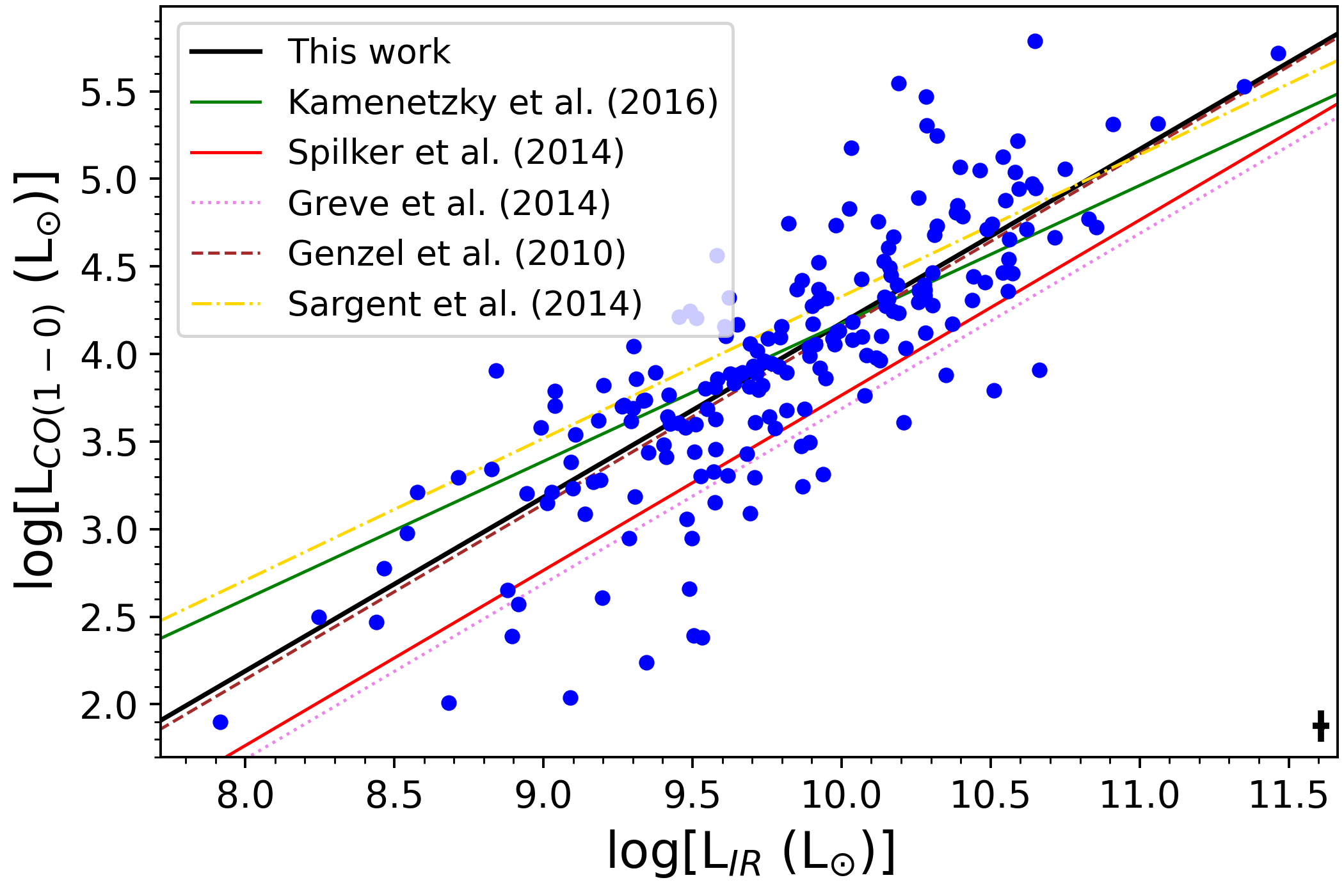}
\caption{Relationship 
 between $\log \rm L_{CO}$ and $\log \rm L_{\rm IR}$. The black solid line represents the relation given by Equation (\ref{eq:3}). Five relations found in the literature are shown for comparison. In the bottom right corner, we show the median uncertainties of the plotted data points.
}
 \label{fig:corr_5}
\end{figure}

The relation of Equation (\ref{eq:3}) is plotted against the data points in Figure\,\ref{fig:corr_5}, where relations found in the literature are also shown for comparison. Such relations are given in terms of L$^\prime_{\rm CO}$, in units of $[\hbox{K}\, \hbox{km}\,\hbox{s}^{-1}\,\hbox{pc}^2]$. We have the following \cite{Solomon1997, CarilliWalter2013}:
\begin{linenomath}
\begin{equation}
{\rm L}_{\rm CO} = 3.2 \times 10^{-11}\nu_r^3 \hbox{L}^\prime_{\rm CO},
\label{eq:Solomon}
\end{equation}
\end{linenomath}
where the rest-frame frequency of the line, $\nu_r$, is in GHz. For the CO(1-0) line, \linebreak  $\nu_r=115.27\,$GHz so that $\log\hbox{L}_{\rm CO}= \log\hbox{L}^\prime_{\rm CO}-4.31$. 

\textls[35]{The relations by \citet{Genzel2010} and by \citet{Kamenetzky2016} refer to the FIR luminosity between 50 and $300\,\upmu$m and between 40 and $120\,\upmu$m, respectively. We converted them to L$_{\rm IR}$ using L$_{\rm IR}(8\text{--}1000\,\upmu{\rm m})$/L$_{\rm FIR}(50\text{--}300\,\upmu{\rm m})=1.3$ and \linebreak  \mbox{L$_{\rm IR}(8\text{--}1000\,\upmu{\rm m})$/L$_{\rm FIR}(40\text{--}120\,\upmu{\rm m})=1.7$} \textls[-15]{as suggested in those papers. \mbox{\citet{Genzel2010}}} derived two L$_{\rm FIR}$--L$^\prime_{\rm CO}$ relations, referring to ``all mergers'' and to ``all SFGs'', respectively. In Figure\,\ref{fig:corr_5}, we plot the ``all SFGs'' one.}

Our best-fit relationship is consistent with a linear relationship, in agreement with the results of \citet{Genzel2010}, \citet{Greve2014}, and \citet{Spilker2014}, based on samples at different redshifts and covering much higher luminosity ranges. Sub-linear relationships were found by \citet{Kamenetzky2016} and \citet{Sargent2014}. Our normalization is in close agreement with that of \citet{Genzel2010} and somewhat higher, although consistent within the errors, with those of \citet{Greve2014} and \citet{Spilker2014}.

Finally, we derived the relationship between $\log \rm SFR$ and $\log \rm L_{\rm CO}$ for our sample using the same methodology. We obtained the following: 
\begin{linenomath}
\begin{equation}
\log (\rm L_{\rm CO}/L_{\odot}) = (3.91\pm0.04) + (0.97\pm0.07) \times \log (\rm SFR/M_{\odot}yr^{-1}).
\label{eq:5}
\end{equation}
\end{linenomath}

\begin{figure}[H] 
\text{~}\\	
\vspace{-10pt}	

\includegraphics[width=0.99\textwidth]{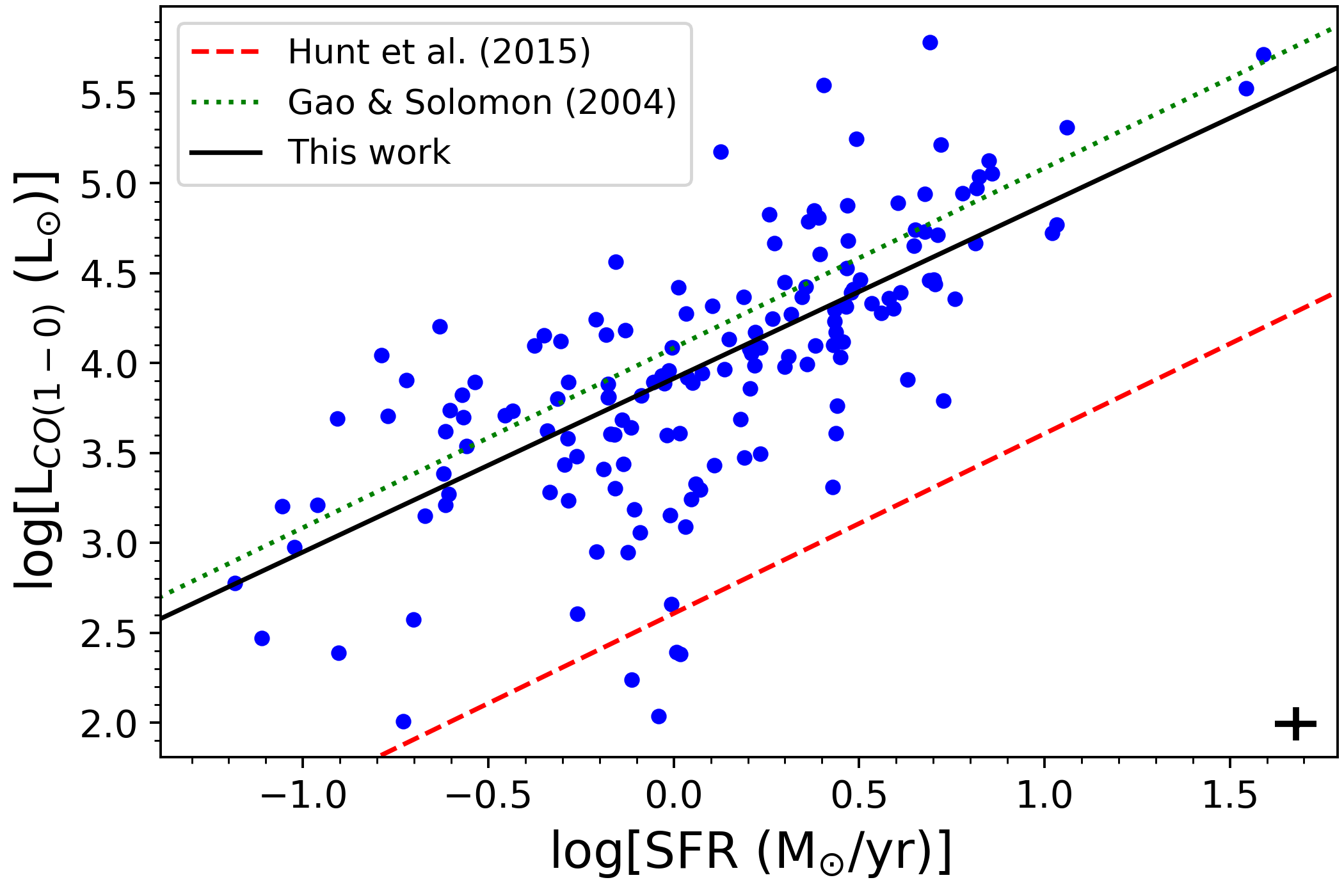}
\caption{Relationship between $\log \rm L_{CO}$ and $\log \rm SFR$. The black solid line represents our relation defined by Equation (\ref{eq:5}). Shown also, for comparison, are the relationships by \citet{GaoSolomon2004} and by \citet{Hunt2015}. The \citet{GaoSolomon2004} relationship was obtained using a sample including normal, luminous, and ultra-luminous infrared galaxies, while the sample of \citet{Hunt2015} is made of metal-poor dwarf galaxies. The cross in the bottom-right corner represents the median uncertainties of the data points.}  
 \label{fig:corr_6}
\end{figure}

The robust confidence intervals at a 95\% confidence level for the intercept and the slope are [3.81,4.02] and [0.77,1.16], respectively.
This relationship is shown in Figure\,\ref{fig:corr_6}, where those by \citet{Hunt2015} and \citet{GaoSolomon2004} are also shown for comparison. 

\textls[-15]{The sample used by \citet{GaoSolomon2004} includes 9 ultra-luminous, \mbox{22 luminous}} infrared galaxies (ULIRGs and LIRGs, respectively), and 34 normal spiral galaxies. The SFR was derived from the IR luminosity. \citet{Hunt2015} measured the CO(1--0) 
 luminosity of eight~metal-poor dwarf galaxies; the SFR rate was computed from the H$\alpha$ and $24\,\upmu$m luminosities. In both cases, the slope of the relation is consistent with unity, but metal-poor dwarf galaxies were found to have a lower CO luminosity at a given SFR by a factor of about 30. Our relationship is in good agreement, within the errors, with the one by Gao and Solomon.

\section{Conclusions}\label{sec:conclusions}
We investigated the correlations between the total (8--$1000\,\upmu$m) IR luminosity, the SFR, and the CO(1-0) luminosity for the DustPedia dataset of local galaxies defined by \citet{Nersesian2019}, removing the 23 galaxies showing either the presence of an AGN or a substantial non-thermal radio contribution to the IR luminosity. Furthermore, we limited our analysis to galaxies whose measurements of the L$_{\rm IR}$ and of the SFR have a signal-to-error ratio of at least three. Our final sample comprises 332 late-type and 56~early-type galaxies, i.e., a total of 388 sources. 

\textls[-25]{The IR luminosity reaches $\log(\hbox{L}_{\rm IR}/\hbox{L}_\odot)\simlt 6.5$ and the SFR reaches $\log(\hbox{SFR}/\hbox{M}_\odot\,\hbox{yr}^{-1}) \simlt -3$}, i.e., values at which L$_{\rm IR}$ is no longer expected to be a good proxy of the SFR. Nevertheless, the data show a statistically significant correlation between the two quantities down to the lowest luminosities and SFRs. 

The relationship between the SFR and L$_{\rm IR}$ for our sample agrees with that by \mbox{\citet{Kennicutt1998}}, which applies to complete absorption by dust of the light emitted by young stellar populations, above $\log(\hbox{L}_{\rm IR}/\hbox{L}_\odot)\simeq 10$ or $\hbox{SFR}\simeq 1\,\hbox{M}_\odot\,\hbox{yr}^{-1}$. The slope is, however, significantly flatter, so our relation substantially exceeds the Kennicutt \cite{Kennicutt1998} one at low luminosity, implying that most of the starlight from young stars is not dust-absorbed. This is at variance with the results by \citet{Clemens2013}, based on a local sample drawn from the \textit{Planck} ERCSC, which implies a dominant role, at low IR luminosity, of dust heating by old stars. Possible origins of the difference are pointed out.

Our sample allowed us to extend the relations between the CO(1-0) luminosity and the L$_{\rm IR}$ or the SFR to low luminosities. Their slopes and their normalizations turned out to be consistent with those obtained at much higher luminosities. 


\vspace{6pt}

\supplementary{A table with classification, L$_{\rm IR}$, SFR, S$_{\rm CO(1\text{-}0)}$, and L$_{\rm CO}$ measurements for the sources of our final sample is provided as online Supplementary Material.
}

\authorcontributions{All 
 authors have contributed substantially to the work~reported. All authors have read and agreed to the published version of the manuscript.}

\funding{M.B. and I.B. acknowledge support from INAF under the mini-grant ``A systematic search for ultra-bright high-z strongly lensed galaxies in Planck catalogues''. V.C. and S.B. acknowledge support from INAF under the mini-grant ``Face-to-Face with the Local Universe: ISM’s Empowerment''.}

\dataavailability{The key data required to reproduce our analysis can be found in the online Supplementary Material.
}


\conflictsofinterest{The authors declare no conflicts of interest.} 

\vspace{6pt} 

\begin{adjustwidth}{-\extralength}{0cm} 
\printendnotes[custom]

\reftitle{References}

\PublishersNote{}
\end{adjustwidth}
\end{document}